\documentclass[12pt,nofootinbib]{revtex4}

\usepackage{amsmath}
\usepackage{amsfonts}
\usepackage{subfigure}
\usepackage{graphicx}
\usepackage{psfrag}
\usepackage{tikz}

\newcommand{\eref}[1]{Eq.~(\ref{#1})}
\newcommand{\nt}{\noindent}
\newcommand{\nn}{\nonumber}
\newcommand{\tr}{\mathrm{tr}}

\def\l{\lambda}

\newcommand{\cO}{{\cal O}}
\def\be{\begin{equation}}
\def\ee{\end{equation}}

\def\bea{\begin{eqnarray}}
\def\eea{\end{eqnarray}}
\def\del{\partial}
\ifx\du\undefined
  \newlength{\du}
\fi
\begin{document}

\title{Dissolved deconfinement: Phase Structure of large $N$ gauge theories with fundamental matter}

\author{Pallab Basu}
 \email{pallab@phas.ubc.ca}
 
 \author{Anindya Mukherjee}
 \email{anindya@phas.ubc.ca}
 
 \affiliation{Department of Physics and Astronomy, 
 University of British Columbia,
 6224 Agricultural Road,
 Vancouver, B.C. V6T 1Z1,
 Canada}

\begin{abstract}
A class of large $N$ $U(N)$ gauge theories on a compact manifold $S^3 \times \mathbb{R}$ (with possible inclusion of adjoint matter) is known to show first order deconfinement transition at the deconfinement temperature. This includes the familiar example of pure $YM$ theory and ${\cal N}=4$ SYM theory. Here we study the effect of introduction of $N_f$ fundamental matter fields in the phase diagram of the above mentioned gauge theories at small coupling and in the limit of large $N$ and finite $N_f/N$. We find some interesting features like the termination of the line of first order deconfinement phase transition at a critical point as the ratio $N_f/N$ is increased and absence of deconfinement transition thereafter (there is only a smooth crossover). This result may have some implication for QCD, which unlike a pure gauge theory does not show a first order deconfinement transition and only displays a smooth crossover at the transition temperature. 

\end{abstract}

\maketitle

\section{Introduction}

Understanding the effect of flavour degrees of freedom in the phase structure of a gauge theory is an important theoretical challenge. QCD with its rich phase structure is an example of such a theory. The ratio $N_f/N$ of flavour and colour degrees of freedom is $O(1)$ for QCD and consequently flavour degrees of freedom have important effects on the QCD phase diagram. While lattice simulations of pure YM theory \cite{Wingate:2000bb,Karsch:2000kv} shows the existence of a first order deconfinement phase transition, incorporation of flavours is thought to make this transition softer (second order) or non-existant \cite{Stephanov:2004wx}. Consequently, it is believed that in real-world QCD there is no first order deconfinement transition. 

One way to study the finite flavour problems is to look at the idealized situation of large-$N$ gauge theories. Here perturbative calculations are easier due to $\frac{1}{N}$ suppression of non-planer diagrams, and in some cases one may use gauge/gravity duality to perform an analytic study of strongly coupled gauge theories\cite{Maldacena:1997re}. For the zero flavour case, a clear analytic demonstration of deconfinement transition for strongly coupled large $N$ ${\mathcal N}=4$  SYM theory was originally discussed by Witten \cite{Witten:1998zw}, where the deconfinement transition in gauge theory is identified with Hawking-Page \cite{Hawking:1982dh} transition in the dual $AdS_5 \times S^5$ background. The case of weakly coupled large $N$ gauge theory on $S_3 \times \mathbb{R}$ is accessible in ordinary perturbation theory and studied  in \cite{AlvarezGaume:2005fv,Aharony:2006rf,Aharony:2005bq}.\footnote{Deconfinement transition in various lower dimensional model large $N$ gauge theories are studied in \cite{Aharony:2005ew}.} Both at strong and weak coupling there is one saddle point at low temperature which is identified as the thermal AdS space. As we increase the temperature to a nucleation temperature ($T_N$) there will be creation of two new saddle points. One is stable (BBH, big black hole in strong coupling) and one is unstable (SBH, small black hole in strong coupling). As the temperature is increased further at certain point($T_1$), the free energy of the big black hole becomes negative and consequently there is a first order transition from the AdS to BBH space. Black holes have a free energy of ${\cal O}(N^2)$ and their gauge theory dual is a deconfined plasma. Hence, the first order transition from the AdS space to the black hole space corresponds to the deconfinement transition in the gauge theory. 

The effects of inclusion of flavours on deconfinement are less well studied. In the limit ($N_f \ll N$) the flavour degrees of freedom will not have significant effects on the phase diagram of the theory and to study the flavour effects we should look at the situations where ${\mathcal O}(N_f/N) \sim 1$. Flavors can be introduced in the context of AdS/CFT by introducing D7 or D5 branes in the D3 brane background \cite{Karch:2002sh}. Although ${\mathcal O}(N_f/N) \sim 1$ means that we can not neglect the back reaction of the flavour branes and one is forced to look at the difficult problem of brane backreaction. Although the exact D3/D7 and other brane intersection solutions has been constructed in the supersymmetric case\cite{Burrington:2004id,Kirsch:2005uy,Cherkis:2002ir}, little is known about the thermal phases and non-extremal solutions \cite{GomezReino:2004pw}. 

\begin{figure}
\begin{center}
\psfrag{ dc}{$\frac{N^c_f}{N}$}
\psfrag{T}{$T$}
\psfrag{ d}{$\frac{N_f}{N}$}
\psfrag{T1}{$T_1$}
\psfrag{Tc}{$T_c$}
\includegraphics[width=5.5cm]{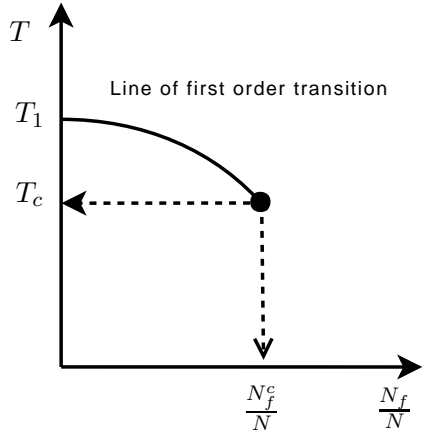}
\end{center}
\caption{Dissolved deconfinement: The line of first order transitions ends at a critical point for nonzero $N_f/N$.}
\label{Fig1}
\end{figure}

In this paper we take the preliminary step to study the flavour effects\footnote{Previous study on similar theories has focused on Gross-Witten-Wadia(GWW) transition \cite{Schnitzer:2004qt,Schnitzer:2006xz} in the finite $N_f/N$ limit. We will also briefly comment on the GWW (\cite{Gross:1980he,Wadia:1979vk,Wadia:1980cp}) transition in the appropriate place.} by introducing flavours in gauge theories defined on compact spaces (e.g., $S^3$). The gauge theories we consider show first order deconfinement transitions in pure form (i.e., without any flavours). We demonstrate that increasing $N_f$ makes this transition softer and eventually for $N_f$ greater than some critical value ($N^c_f$) there will be no deconfinement transition (see Fig \ref{Fig1}). We also comment on the effect of the flavour mass on the phase diagram. As expected it is found that flavours gradually get decoupled if we increase the flavour mass. This result seems to have some implication for real world QCD and as we discuss may possibly be continuously connected to that case.


\subsection{Effective action and order parameter}

We first present a general discussion of the effective action of a generic weakly coupled $SU(N)$ YM theory on the compact manifold $S^3$. We consider the theory in the canonical ensemble, i.e., the Euclidean time direction is periodically identified with a period of $\beta = {1 \over T}$. It was shown in \cite{Sundborg:1999ue,Aharony:2003sx} that the Yang-Mills theory partition function on $S^3$ at a temperature $T$ can be reduced to an integral over a unitary $SU(N)$ matrix $U$, which is the zero mode of the Polyakov loop on the Euclidean time circle. Their analysis was done in the limit when the 't Hooft coupling $\lambda << 1$.
 \begin{equation}
  Z (\lambda,T)= \int dU \, e^{-S (U)}
  \end{equation}
with
 \begin{equation}
 U = P \exp \Bigl(i \int_0^{ \beta} A_{0} d \tau \Bigr)
 \end{equation}
where $A_{0} (\tau) $ is the zero mode of the time component of the gauge field on $S^3$. This follows from the fact that apart from $A_{0}$ all modes of the gauge theory on $S^3$ are massive and can be integrated out. Hence we can use $U$ as an order parameter. 

 In the $\lambda=0$ case the the partition function of a generic gauge theory defined on a compact manifold is presented in \cite{Aharony:2003sx}. Since the theory is on a compact space, there is a constraint from Gauss' law which causes interaction between the gauge and other degrees of freedom. The effective action is,
\begin{equation}
 \label{genform}
 \mathbb{Z}(x) = \int [dU] \exp \left\{\sum_{R,n}\frac1n \left[z_B^R(x^n)+(-1)^{n+1}z_F^R(x^n)\right]\cdot\chi_R(U^n)\right\},
\end{equation}
where $U$ is a $N\times N$ unitary matrix, $\chi_R(U)$ is the character for the representation $R$, $z_B^R(x)$ and $z_F^R(x)$ are the single particle bosonic and fermionic partition functions for the representation $R$ (\cite{Aharony:2003sx}):
\begin{equation}
 \label{singlep}
 z_B^R(x) = \sum_{R_i = R} x^{E_i}, \qquad z_B^R(x) = \sum_{R_i = R} x^{E'_i},
\end{equation}
where $E_i, E'_i$ are respectively the energies of the bosonic and fermionic states. For adjoint or fundamental matter fields on $S^3 \times \mathbb{R}$ this is given by:

\begin{equation}
z_{B}(x) = {x+ x^2 \over (1-x)^3} , \qquad z_{F}(x) = {4 x^{3 \over 2} \over (1-x)^3}
\label{behav}
\end{equation}

The quantity $x$ is related to the temperature $T$ by $x = e^{-1/T}$. We are interested in a $SU(N)$ gauge theory on $S^3 \times \mathbb{R}$ with $N_f$ matter fields in the fundamental representation, and $N_f/N$ finite. Let $z_B(x)$ and $z_F(x)$ be the single particle partition functions for the adjoint representation, and $Z_B(x)$ and $Z_F(x)$ that for the fundamental representation (note that $z(x)$ and $\mathcal{Z}(x)$ are similar in form, both being defined by the general \eref{singlep}). Hence we have from \eref{genform}:
\begin{eqnarray}
  \label{actualform}
  \mathbb{Z}(x) &=& \int [dU] \exp\left\{\sum_{n=1}^\infty \frac1n [\mathcal{Z}_B(x^n)+(-1)^{n+1}\mathcal{Z}_F(x^n)] \left(\tr(U^n) + \tr(U^{-n}) \right)\right. \nn \\
  &+& \left.\sum_{n=1}^\infty \frac1n \left[z_B(x^n)+(-1)^{n+1}z_F(x^n)\right] \tr(U^n)\tr(U^{-n})\right\}.
\end{eqnarray}

Now consider the same theory with non-vanishing 't~Hooft coupling $\l$.  Gauge invariance requires that the effective action of $U$ be expressed in terms of products of $\tr\,U^n$, with $n$ an integer, since these are the only gauge invariant quantities that can be constructed from $A_{0}$ alone. It is useful in this section to keep any scalar self-couplings and Yukawa couplings at zero. The partition function can be written as \cite{Aharony:2003sx,AlvarezGaume:2006jg,Schnitzer:2006xz}
 \begin{equation}
 Z\!\!\!Z (x) = \int [dU] \exp \left[ -
 \tilde{S}_{eff} (\rho_n)\right] \; .
 \end{equation}
where $\rho_n=\frac{\tr(U^n)}{N}$. 
The planar contribution at large $N$ to $S_{eff}$ in perturbation
theory is, in a double expansion in $\l$ and $N_f/N$, consistent with
gauge invariance:

\begin{eqnarray}
 N^{-2} \tilde{S}_{eff} (\rho_n) & = & F^{0}(\rho_j,\lambda,\beta) \\
\nonumber & + & (\frac{N_f}{N}) \sum_{n} F^1_{n}(\rho_j,\lambda,\beta) \rho_n  + \l (\frac{N_f}{N})^2 \sum_{n_1,n_2} F^{2}_{n_1,n_2}(\rho_j,\lambda,\beta) \rho_{n_1} \rho_{n_2}\\ 
\nonumber & + & \cdots + \l^{p-1}(\frac{N_f}{N})^p \sum_{n_1,...,n_p} F^{p}_{n_1,\cdots,n_p}(\rho_j,\lambda,\beta) \prod_{i=1}^{i=p} \rho_{n_i} +\cdots
\label{effective}
\end{eqnarray}
where, each function $F^{p}_{n_1,\cdots,n_p}(\rho_j,\lambda,\beta)$ appearing in above formula is built up by a series of $Z_N$ invariant product of $\rho_n$'s, \footnote{In general the coefficients $a_s{r_1,r_2,...r_k}^{m_1,m_2,...m_k}(\lambda,\beta)$ is different for different $F^{p}_{n_1,\cdots,n_p}(\rho_j,\lambda,\beta)$. However for simplicity we have not shown this explicitly.}
\begin{eqnarray}
 F^{p}_{n_1,\cdots,n_p}(\rho_j,\lambda,\beta)=\sum_{\sum_{i=1}^{k} r_i m_i=0} a_{r_1,r_2,...r_k}^{m_1,m_2,...m_k}(\lambda,\beta) \prod_{i=1}^{k} \rho_{r_i}^{m_i} 
\end{eqnarray}
In perturbation theory we have, $a_{r_1,r_2,...r_k}^{m_1,m_2,...m_k}(\lambda,\beta)=\lambda^{\sum_i |r_i m_i|}f_{0}(\beta)$+higher order terms in $\lambda$. 

 The form of effective action (\ref{effective}) is determined by considering the Feynman diagrams of the theory, where each insertion of a fundamental loop into an adjoint line of a diagram gives a
factor of $\l (N_f/N)$, while a radiative correction by an adjoint
line to a fundamental propagator gives a factor of $\l$ \cite{Veneziano:1976wm} (Since we are keeping scalar self-couplings and Yukawa couplings at zero,
these are the only corrections to be considered in this section.).

In perturbation theory, near the saddle point of \eref{effective} one may integrate out the $\rho_n
(n\geq 2)$\footnote{By doing so one loose any information about the multigap phases \cite{Bhanot:1990bd}, but this here we are not interested in that aspect.}, since at near the  interesting phase transition regime,
$|\rho_n (x_c)|<<|\rho_1 (x_c)|$ for $n \geq  2$ (see \cite{Aharony:2005bq,Schnitzer:2006xz} for a detailed discussion.). Therefore the effective action for $\rho_1$ near the saddle-point for $\rho_1$ is
 \bea
\label{flaveffac}
 {N^{-2} \tilde{S}_{eff} (\rho_1)} & = & \left[ -m_1^2 |\rho_1 |^2 + b |\rho_1|^4
 + \ldots \right]  \nonumber
 \\
 & + & \left( \frac{N_f}{N}\right) \Big\{ (\rho_1 + \bar{\rho}_1) \left[
 -c + e |\rho_1|^2 + f |\rho_1|^4 + \ldots \right]  \nonumber
 \\
&+& {\cal O}(\lambda^3)+{\cal O}\left({N_f \over N}^2\right)
 \eea
The leading order contributions to the coefficients in (2.6) are
 \bea
 m_1^2 & = & {\cal O} (1) \; \; ; \;\; b \; = \; \cO (\l^2 ) \;\; ;
 \nonumber \\
 c & = & {\cal O} (1)  {\rm \; and \; positive \; at \; leading \; order}
 \;\; ; \nonumber \\
 e & = & {\cal O} (\l^2 ) \; \; ; \;\; f \; = \; {\cal O} (\l^4 )
 \nonumber
 \eea
From \eref{actualform} we can read off the temperature dependence of $m_1^2$ and $c$:
\begin{eqnarray}
 \label{m1sq}
 m_1^2 &=& z_B(x) + z_F(x). \\
\nonumber c &=& \mathcal{Z}_B(x) + \mathcal{Z}_F(x) 
\end{eqnarray}
Since each insertion of a fundamental loop in an adjoint line is
accompanied by a factor of $\l (N_f/N)$, higher-orders in $N_f/N$ result in
higher powers of $\l$ compared to those displayed in (\ref{flaveffac}), while
the ${\cal O} (N_f/N)$ term comes from the fundamental loop.

A large $N$ unitary matrix model such as \eref{effective} can be solved using the saddle point method by introducing a density variable $\rho(\theta)={1 \over N} \sum \delta(\theta-\theta_i)$, where $e^{\theta_i}$'s are the eigenvalues of the unitary matrix. We have,
\begin{eqnarray}
 \int_{-\pi}^\pi \rho(\theta) d\theta = 1. \\
\rho(\theta)>0
\end{eqnarray} 
The terms of $\rho_n$, $\rho(\theta)$ can be written as:
\begin{equation}
 \rho(\theta) = \frac1{2\pi} \sum_{n=-\infty}^\infty \rho_n e^{in\theta}.
\end{equation}
To solve the theory we also have to take into account the measure contribution ($S_M$) in the total action($S_{\mathrm{tot}}$). At large N we have,
\begin{equation}
 S_{\mathrm{tot}}([\rho])=S_{\mathrm{eff}}([\rho])+S_M([\rho])
\end{equation}
where $S_M([\rho])=\int d\theta d\phi \log(\cos(\frac{\theta-\phi}{2})) \rho(\theta)\rho(\phi)$. Written in terms of $\rho_n$, the measure contribution can have different analytic expressions depending on whether $\rho(\theta)$ is gapped or ungapped. For example if we concentrate on an effective action which is a function of a single moment $\rho_1$, then $S_M(\rho_1)$ is given by:
\begin{eqnarray}
 S_M(\rho_1) &=& \rho_1^2, \qquad \rho_1 < \frac12 \nn \\
             &=& \frac14 - \frac12 \ln 2(1-\rho_1), \quad \rho_1 \ge \frac12.
\end{eqnarray} 


\section{No flavour case: Deconfinement transition}
 In this chapter we will review the previous works on deconfinement transition of weakly coupled gauge theory on $S^3 \times \mathbb{R}$. \footnote{Most clear analytic demonstration of deconfinement transition in the strong coupling comes from the Hawking-Page transition in $AdS_5 \times S^5$ \cite{Witten:1998zw}}.  From \eref{flaveffac} we get for $N_f=0$ case,
\begin{equation}
\label{noflav}
 N^{-2}S_\mathrm{eff} = -m_1^2 |\rho_1|^2 - b|\rho_1|^4 
\end{equation}
Taking into account the measure factor for $U(N)$, the saddle point equation for $\rho_1$ is:
\begin{eqnarray}
 \label{saddle}
 m_1^2 \rho_1 + 2b \rho_1^3 &=& \rho_1, \qquad \rho_1 < \frac12 \nn \\
 &=& \frac1{4(1-\rho_1)}, \quad \rho_1 \ge \frac12.
\end{eqnarray}
The free energy at a saddle is given by:
\begin{eqnarray}
\label{free}
 N^{-2}S_\mathrm{tot} &=& -m_1^2 \rho_1^2 - b \rho_1^4 + 2d \rho_1 + \rho_1^2, \qquad \rho_1 < \frac12 \nn \\
   &=& -m_1^2 \rho_1^2 - b \rho_1^4 + 2d \rho_1 - \frac12 \ln 2(1-\rho_1) + \frac14, \quad \rho_1 \ge \frac12.
\end{eqnarray} 
Depending on the sign of $b(\beta,\lambda)$ the theory may exhibit different phase structures. It has been shown in \cite{Aharony:2005bq}, that $b>0$ for the perturbative pure YM theory on $S^3$ and it is expected that same holds for the ${\cal N}=4$ SYM theory also. If $b>0$ the phase structure of the weakly coupled gauge theory resembles{\cite{AlvarezGaume:2005fv}} that of strongly coupled ${\cal N}=4$ SYM theory derived from $AdS/CFT$ and also that of strongly coupled pure large $N$ YM theory. It seems possible that the phase diagram can be extrapolated to strong coupling. In the next section we will review the phase diagram for the various values of $b(\beta,\lambda)$.

\subsection{$N_f/N = 0,~\lambda = 0$}

\begin{figure}[h!]
 \begin{center}
   \psfrag{xlabel}{$\rho_1$}
   \psfrag{ylabel}{$\frac1{N^2} S_\mathrm{tot}(\rho_1)$}
   \psfrag{zlabel}{$\frac1{N^2}\frac{\del S_\mathrm{tot}}{\del \rho_1}$}
   \psfrag{key1}[r][r][0.5]{$T < T_H$}
   \psfrag{key2}[r][r][0.5]{$T = T_H$}
   \psfrag{key3}[r][r][0.5]{$T > T_H$}
   \subfigure[$\frac1{N^2}\frac{\del S_\mathrm{tot}}{\del \rho_1}$]{\label{fig:purehagea} \includegraphics[width=5.5cm]{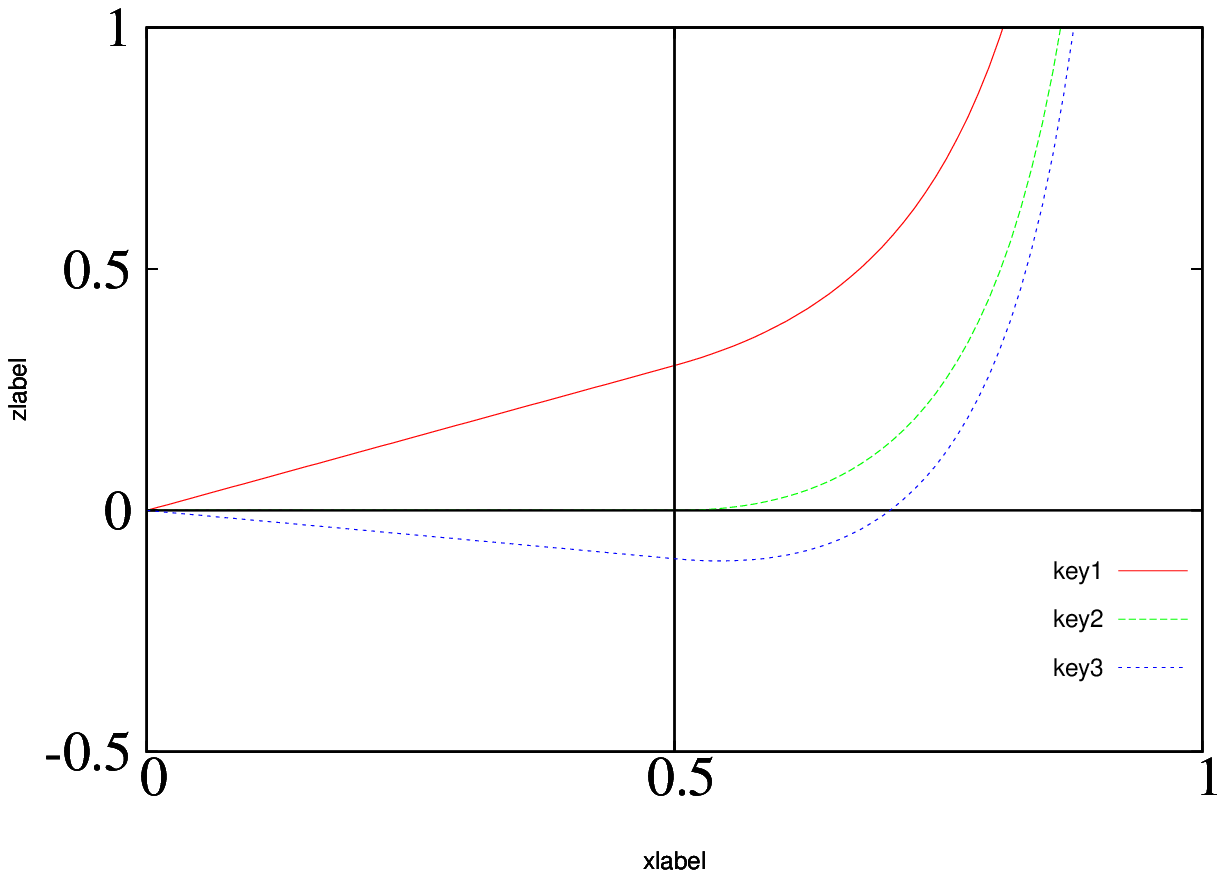}} \hspace{0.1cm}
   \subfigure[$\frac1{N^2} S_\mathrm{tot}(\rho_1)$]{\label{fig:purehageb} \includegraphics[width=5.5cm]{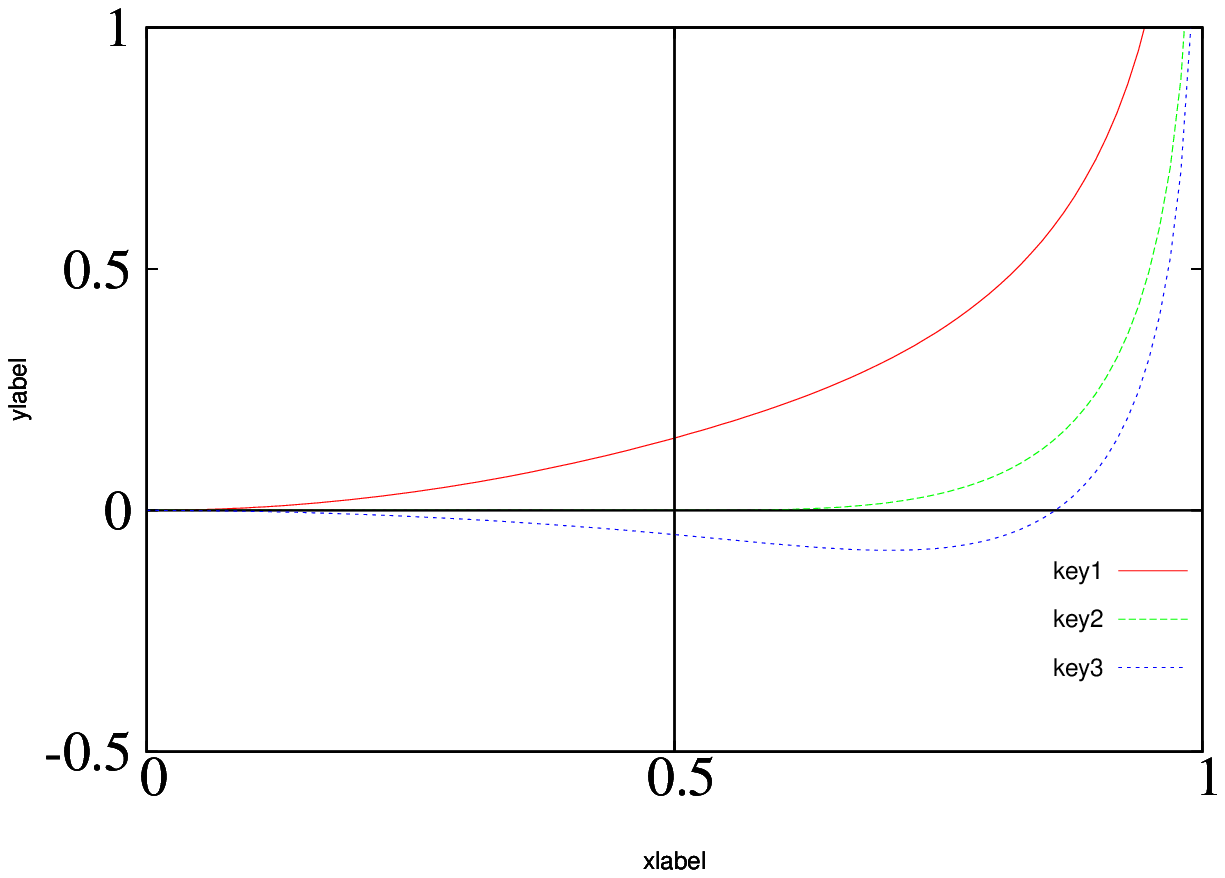}}
 \end{center}
 \caption{First order phase transition at $N_f/N = 0,~\lambda = 0$.}
 \label{fig:purehage}
\end{figure}

This corresponds to $b=0,~d=0$. In this case only the temperature is variable. From \eref{saddle} we see that there is always a solution for $\rho_1 < \frac12$ at $\rho_1 = 0$. 
This solution has free energy of order O(1). For $m_1^2<1$ there is no other solution. For $m_1^2 > 1$ there is another solution with $\rho_1 \ge \frac12$ which comes from solving from the second line of \eref{saddle}. This solution is:
\begin{equation}
\label{heg}
 \rho_1^{II} = \frac{1 + \sqrt{1-\frac{1}{m_1^2}}}2.
\end{equation}
   There is no change in this picture as $m_1^2$ is gradually increased until we reach $m_1^2 = 1$. At Hagedorn temperature $T_H$, $m_1^2(T_H) = 1$ and a flat direction is created around $\rho_1 = 0$. As we increase temperature further ($T > T_H$) we have $m_1^2 > 1$ and the dominant saddle point of the system jumps from $\rho_1=0$ to $\rho_1=\rho^{II}_1$. The free energy of the new saddle point is negative and is of ${\cal O}(N^2)$. This is the Hagedorn transition in a free theory \cite{Pisarski:1983db,Sundborg:1999ue,Aharony:2003sx}. The flat direction at $m_1^2 = 1$ is an artifact of the free theory and is lifted by the introduction of a small coupling. Fig \ref{fig:purehagea} shows a plot of the LHS of the saddle point equation, $\frac1{N^2} \frac{\del S_\mathrm{tot}}{\del \rho_1}$ vs. $\rho_1$. A new saddle point is created when this curve crosses the $\rho_1 = 0$ line. Fig \ref{fig:purehageb} shows the corresponding free energy $N^{-2}S_\mathrm{tot}(\rho_1)$.

\subsection{$N_f/N = 0,~\lambda \ne 0$}

\subsubsection{$b > 0$}

\begin{figure}[h!]
 \begin{center}
   \psfrag{xlabel}{$\rho_1$}
   \psfrag{ylabel}{$\frac1{N^2} S_\mathrm{tot}(\rho_1)$}
   \psfrag{zlabel}{$\frac1{N^2}\frac{\del S_\mathrm{tot}}{\del \rho_1}$}
   \psfrag{key1}[r][r][0.5]{$T < T_N$}
   \psfrag{key2}[r][r][0.5]{$T = T_N$}
   \psfrag{key3}[r][r][0.5]{$T > T_c$}
   \subfigure[$\frac1{N^2}\frac{\del S_\mathrm{tot}}{\del \rho_1}$]{\label{fig:firsta} \includegraphics[width=5.5cm]{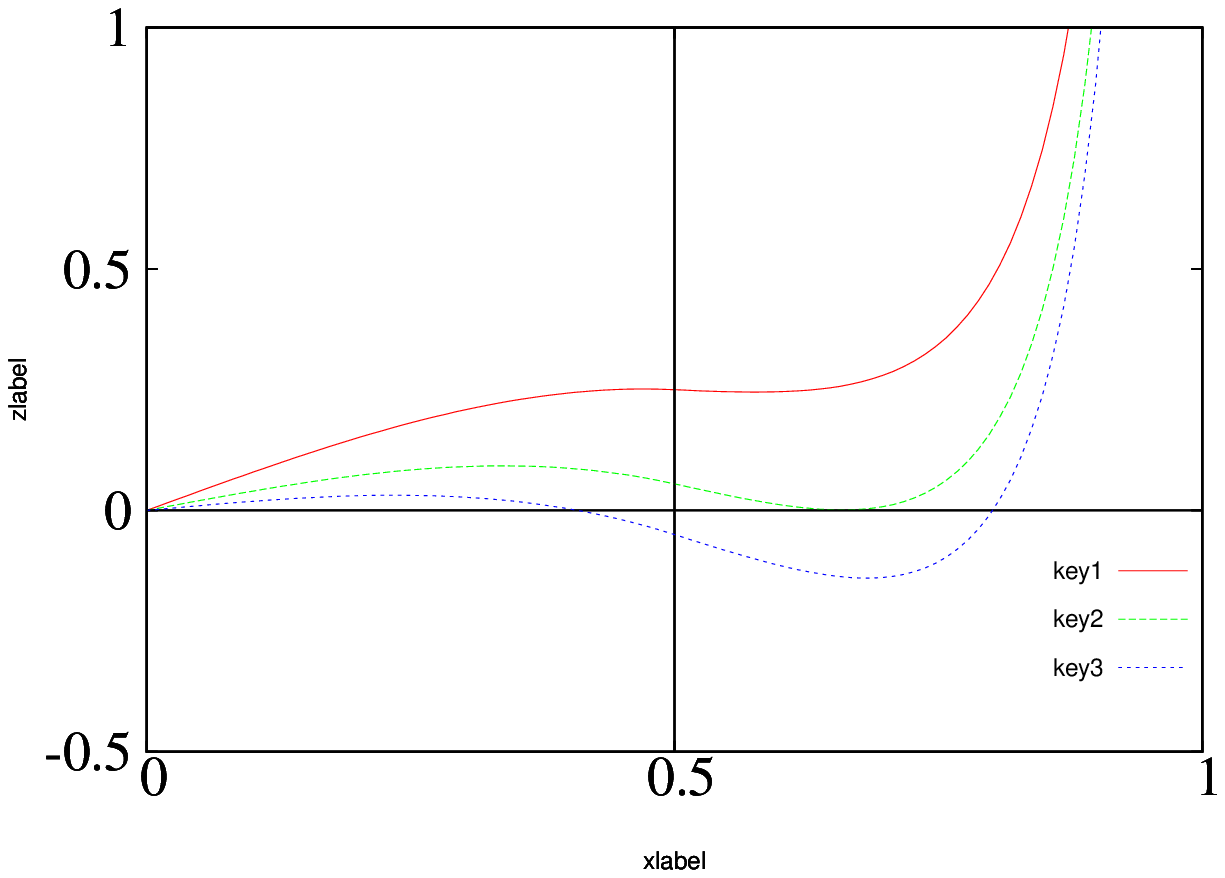}} \hspace{0.1cm}
   \subfigure[$\frac1{N^2} S_\mathrm{tot}(\rho_1)$]{\label{fig:firstb} \includegraphics[width=5.5cm]{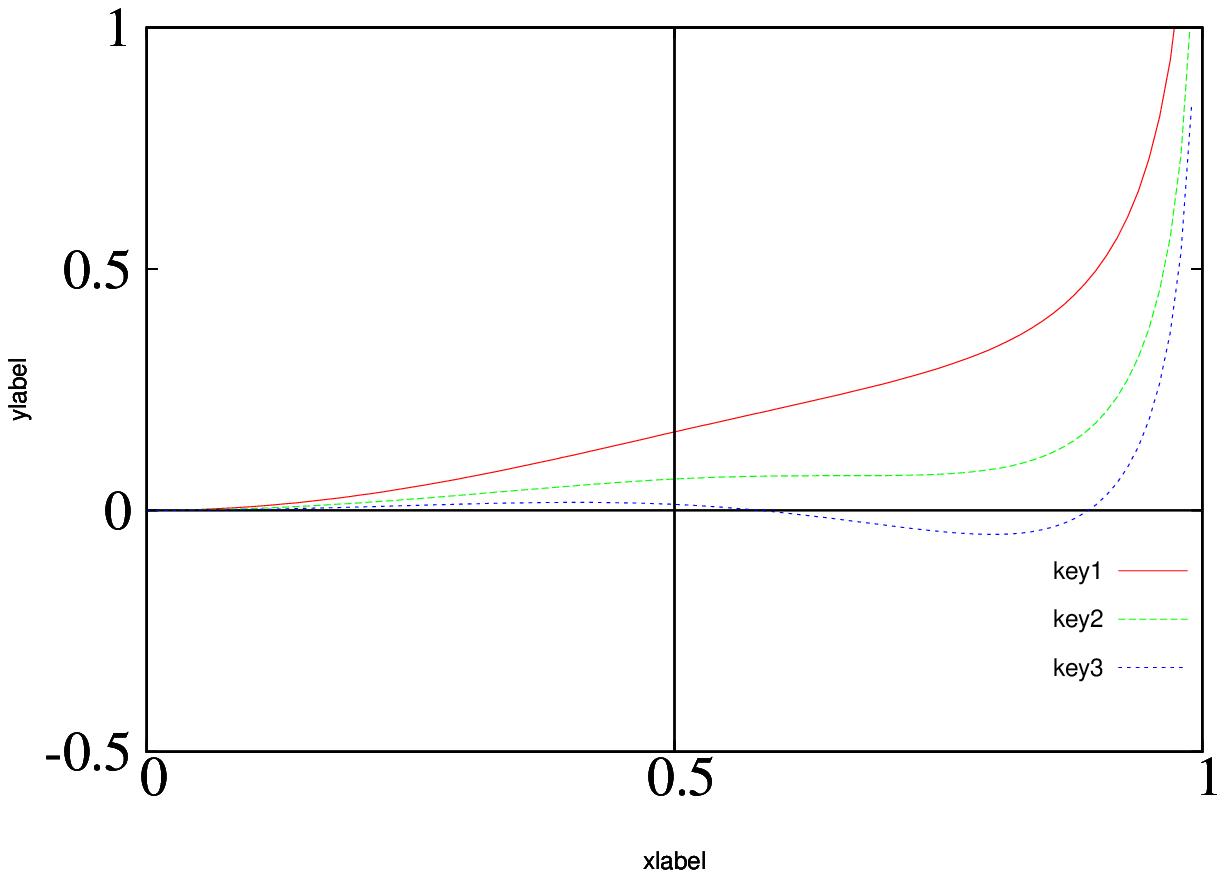}}
 \end{center}
 \caption{First order phase transition at $N_f/N = 0,~\lambda \ne 0$, $b$ positive.}
 \label{fig:first}
\end{figure}

Next we consider the case when a small coupling is turned on. This sets $b \ne 0$ in \eref{saddle}. The nature of the transition in this case depends on the sign of the coefficient $b$. Let us first take $b$ to be a small positive value. Then we see from \eref{saddle} that there is always an extremum at $\rho_1 = 0$. For $T < T_H$ this saddle point is stable. As temperature is increased, two things happen: first, two new saddle points appear at a temperature $T_N$ (nucleation temperature) which is below $T_H$ (see Fig \ref{fig:first}). The rightmost saddle is stable, while the middle one is unstable. As temperature is further increased, the rightmost saddle becomes dominant at a temperature $T_c < T_H$ (with a negative free energy), and at this stage the system jumps from the saddle at $\rho_1 = 0$ to the rightmost one. This transition is first order, and the free energy of the system jumps from $O(1)$ to $O(N^2)$. It can be shown using Eqs. (\ref{saddle}), (\ref{free}) that $T_c$ decreases with increasing $\lambda$.

\subsubsection{$b < 0$}

\begin{figure}[h!]
 \begin{center}
   \psfrag{xlabel}{$\rho_1$}
   \psfrag{ylabel}{$\frac1{N^2} S_\mathrm{tot}(\rho_1)$}
   \psfrag{zlabel}{$\frac1{N^2}\frac{\del S_\mathrm{tot}}{\del \rho_1}$}
   \psfrag{key1}[r][r][0.5]{$T < T_H < T_c$}
   \psfrag{key2}[r][r][0.5]{$T_H < T < T_c$}
   \psfrag{key3}[r][r][0.5]{$T > T_c$}
   \subfigure[$\frac1{N^2}\frac{\del S_\mathrm{tot}}{\del \rho_1}$]{\label{fig:first2a} \includegraphics[width=5.5cm]{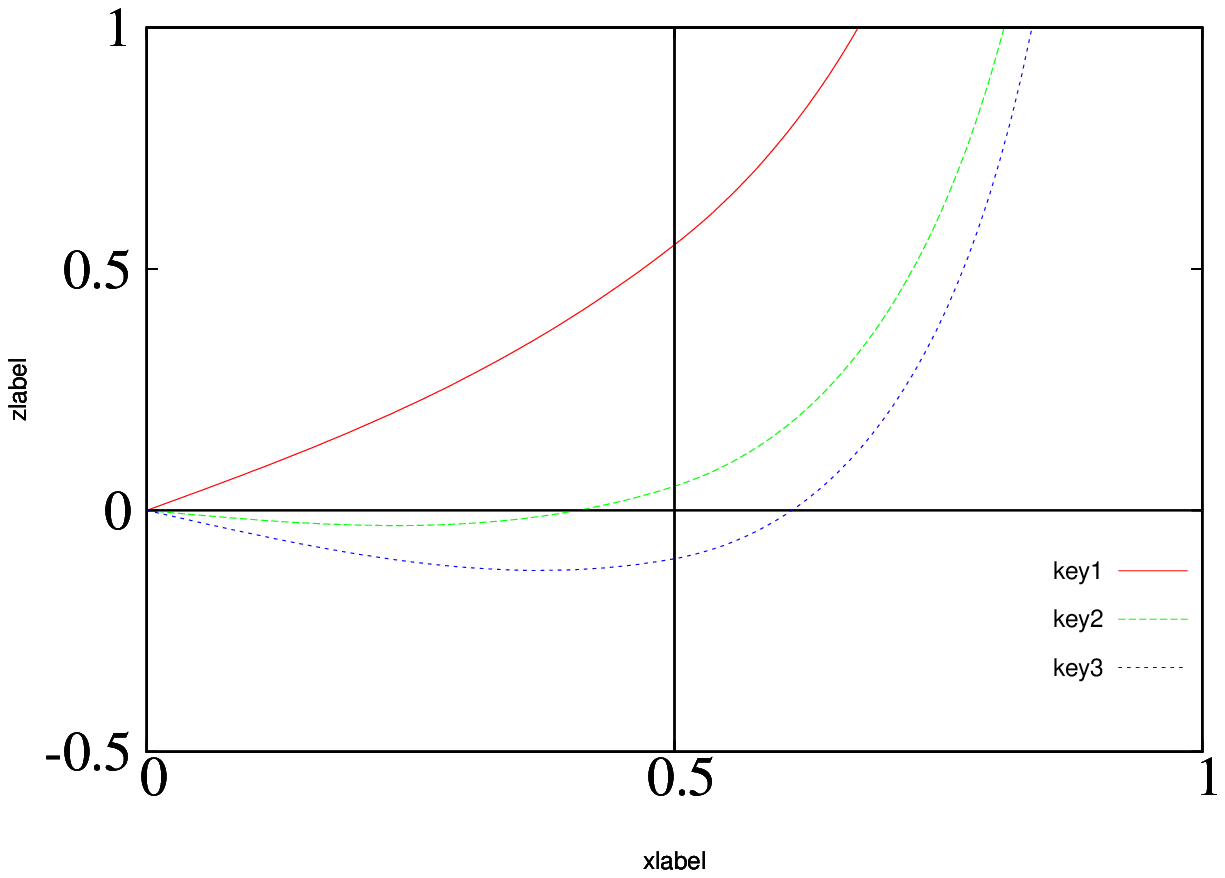}} \hspace{0.1cm}
   \subfigure[$\frac1{N^2} S_\mathrm{tot}(\rho_1)$]{\label{fig:first2b} \includegraphics[width=5.5cm]{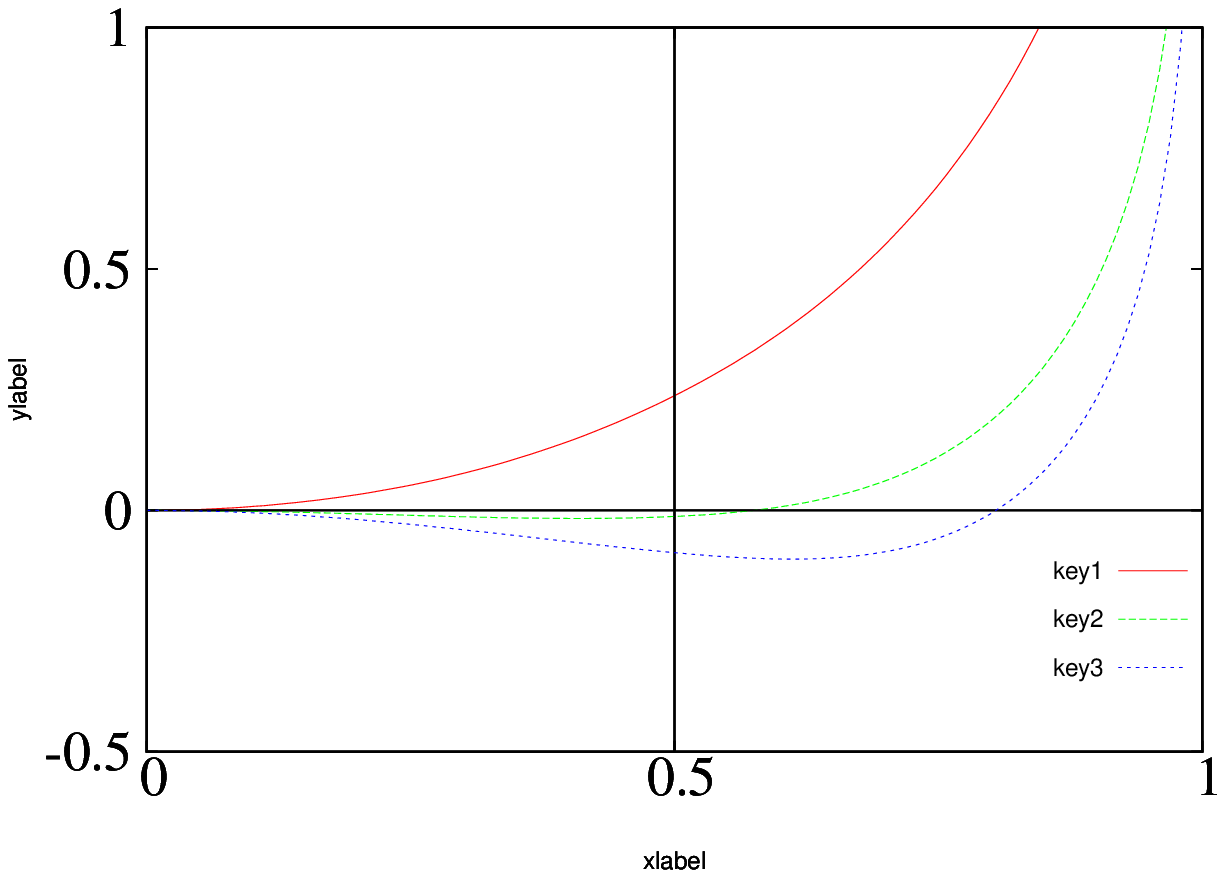}}
 \end{center}
 \caption{Phase structure for $N_f/N = 0,~\lambda \ne 0$, $b$ negative.}
 \label{fig:first2}
\end{figure}

We now consider the situation when $b$ is negative. In this case nothing happens below $T_H$. At the Hagedorn temperature $T_H$ the system undergoes a first order transition much like the case $\lambda = 0$. As the temperature is increased further, the new stable saddle shifts towards increasing $\rho_1$. At a temperature $T_c$ it crosses $\rho_1 = \frac12$. This causes a third order transition, as the third derivative of the free energy \eref{free} is discontinuous at this point (see Fig \ref{fig:first2}).

\section{$N_f/N \ne 0$: Dissolved deconfinement}
For $N_f \ne 0$ the effective action looks like:
\begin{equation}
\label{wthflav}
 N^{-2}S_\mathrm{eff} = -m_1^2 |\rho_1|^2 - b|\rho_1|^4 + d(\rho_1 + \bar{\rho}_1).
\end{equation}
where we have kept only the first terms in $\l$ and $\frac{N_f}{N}$ expansion in \eref{flaveffac} and
\begin{equation}
\label{dval}
d = c \frac{N_f}{N}.
\end{equation}
 The saddle point equations we have to solve in this case are just the simple modifications of \eref{saddle} with an additional term for $d$
\begin{eqnarray}
 \label{saddle2}
 m_1^2 \rho_1 + 2b \rho_1^3 - d &=& \rho_1, \qquad \rho_1 < \frac12 \nn \\
 &=& \frac1{4(1-\rho_1)}, \quad \rho_1 \ge \frac12.
\end{eqnarray}
we will show that the effect of the term $d$ will be to make the deconfinement transition softer and then non-existant. It should be noted that from Eqs. (\ref{m1sq}), (\ref{dval}) \ that $d$ can be tuned both by tuning $T$ and $N_f/N$ and if we increase the mass of the flavours, $d$ becomes small\footnote{For large flavor mass $m$, it follows from \eref{singlep} $d \propto \exp{\frac{-m}{T}}$.} . At first we will discuss the $b = 0$ case and then we will discuss the more interesting $b > 0$ case. 

\subsection{$N_f/N \ne 0,~\lambda = 0$}
\begin{figure}[h!]
 \begin{center}
   \psfrag{xlabel}{$\rho_1$}
   \psfrag{ylabel}{$\frac1{N^2} S_\mathrm{tot}(\rho_1)$}
   \psfrag{zlabel}{$\frac1{N^2}\frac{\del S_\mathrm{tot}}{\del \rho_1}$}
   \psfrag{key1}[r][r][0.5]{$T < T_c$}
   \psfrag{key2}[r][r][0.5]{$T > T_c$}
   \subfigure[$\frac1{N^2}\frac{\del S_\mathrm{tot}}{\del \rho_1}$]{\label{fig:thirdnewa} \includegraphics[width=5.5cm]{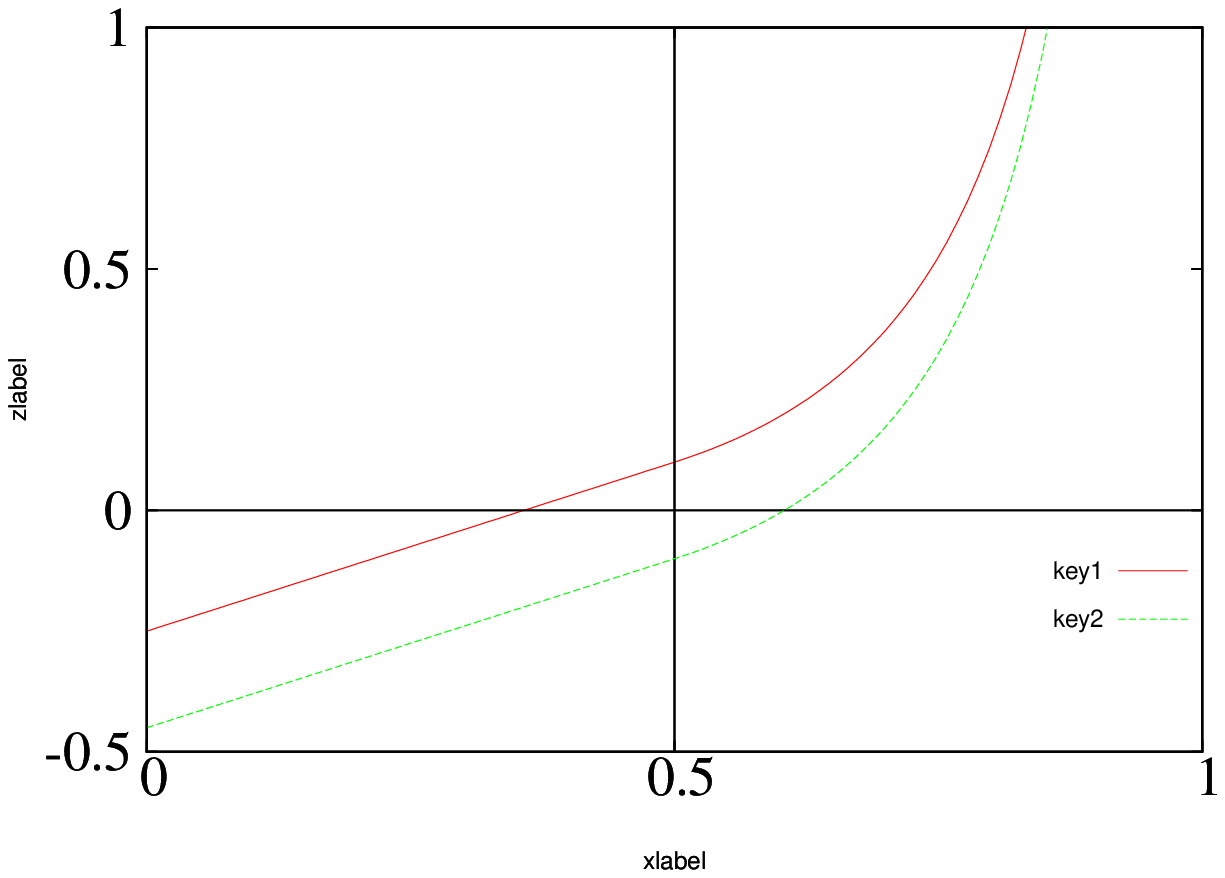}} \hspace{0.1cm}
   \subfigure[$\frac1{N^2} S_\mathrm{tot}(\rho_1)$]{\label{fig:thirdnewb} \includegraphics[width=5.5cm]{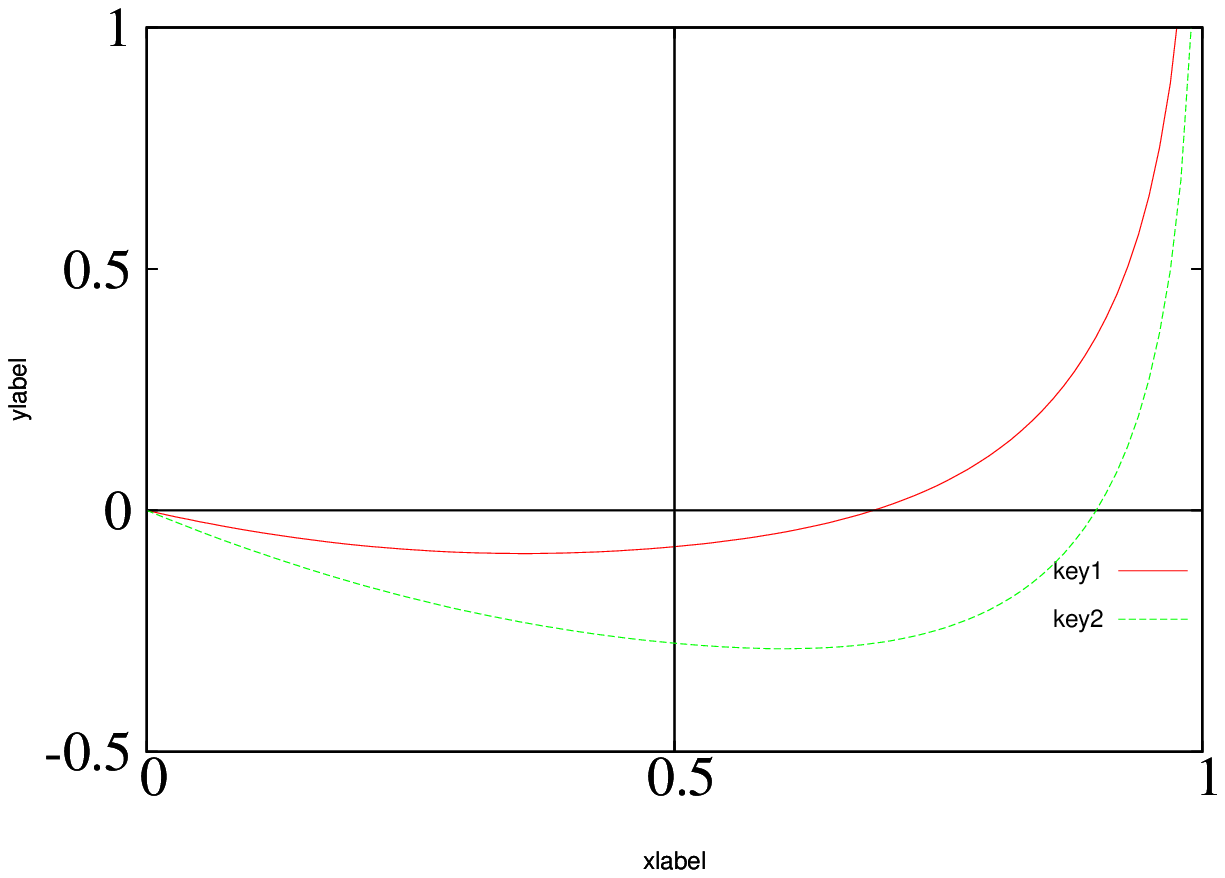}}
 \end{center}
 \caption{Third order phase transition at $N_f/N \ne 0,~\lambda = 0$.}
 \label{fig:thirdnew}
\end{figure}

Here we keep the coupling zero (and hence $b=0$), while increasing the value of $N_f/N$. The quantity $d$ is negative in our conventions. In this case, changing $d$ at a fixed temperature (and hence $m_1$) only changes the offset of \eref{saddle}, and so there is always only one saddle point. As argued in \cite{Schnitzer:2004qt}, this saddle point has $\rho_1 \neq 0$. The position of this saddle point on the $\rho_1$ axis shifts with changing $d$. However, the RHS of \eref{saddle} has a discontinuity in its second derivative (and hence the free energy has a discontinuity in its third derivative) at $\rho_1 = \frac12$. So there is a third order phase transition (GWW transition) when the saddle point crosses $\rho_1 = \frac12$ and the minimum of the free energy goes to $\rho_1 > \frac12$. The temperature at which this happens decreases with increasing $N_f/N$, since increasing temperature and $N_f/N$ both have the effect of pushing the saddle point to the right, towards $\rho_1 = \frac12$, if we start in the phase where $\rho_1 < \frac12$. The value of $m_1^2$ at the transition point is given by:
\begin{equation}
  m_1^2 = 1 + 2d.
\end{equation}
Recall that $d$ is negative in our model. Fig \ref{fig:thirdnew} shows the transition.

 Let us take a closer look at the region near $T=0$. The saddle point for the system is, from \eref{saddle2}
\begin{equation}
 \label{momentsaddle}
 \rho = -d/(1-m_1^2) 
\end{equation} 
Now from \eref{m1sq} it follows that as $T\to 0$, $m_1^2,d \to 0$. So $\rho_n \to 0$ as $T\to 0$. So in this region the eigenvalue distribution $\rho(\theta)$ becomes uniform. This corresponds to a geometry without an event horizon, which tallies with the brane picture.

\subsection{$N_f/N \ne 0,~\lambda \ne 0$}
\label{1orderterm}

Let us now consider the case with nonzero $N_f/N$ and coupling (we also assume $b > 0$). At non-zero $d$ we always have $\rho_1 \ne 0$ at finite temperature. (Recall that according to the discussion of the previous subsection $T \rightarrow 0$ $\Longrightarrow$ $\rho_1 \rightarrow 0$.) In low temperature phase we have one saddle point($I$). In fact one can easily calculate that at low temperature,
\bea
\rho_1 \approx \frac{d}{m_1^2-1} \approx d
\eea 
Here we have used the fact $m_1^2 \rightarrow 0$ as $T \rightarrow 0$. As we have studied numerically, increasing the temperature(same as increasing $m_1^2$) creates two new saddle points. One of them is stable ($III$) and other is unstable ($II$). The saddle points are nucleated in the gapped phase. As we increase $m_1^2$ further, the value of $\rho_1$ at saddle point $III$  increases, whereas the value of $\rho_1$ at saddle point $II$ decreases. At some point the unstable saddle point $II$ crosses $\rho_1=\frac{1}{2}$ and consequently goes through a Gross Wadia Witten (GWW) transformation.

At the nucleation temperature both of the new saddle points have positive free energy. Hence the dominant saddle point of the system is the saddle point $I$. However as we increase the value of $m_1^2$, the free energy difference ($S_{III}-S_{I}$) between the saddle points $III$ and $I$ decreases and at $m_1^2 = m_1^{2,\mathrm{1^{st}}}(d)$ there is a first order transition between saddle point $I$ and saddle point $III$. After this  $III$ becomes the dominant saddle point of the system. This is the analogue of deconfinement transition at finite but small $N_f$. When   $d$ is sufficiently small the free energy at the saddle point $I$ is given by $S_I \propto N^2 d^2 \propto N_f^2$. This free energy comes from the effects of messons whose number is ${\cal O}(N_f^2)$. In contrast, the saddle point $III$ has a free energy of order $N^2$ which comes from the effect of the adjoint (gluonic) degrees of freedom. Hence at small $d$ this first order transition may indeed be thought as a transition from a confined mesonic plasma to a deconfined gluonic plasma. At a still higher value of $m_1^2 = m_1^{2,II\rightarrow I}(d) > m_1^{2,\mathrm{1^{st}}}(d)$, the saddle points $II$ and $I$ meet and after that there is just saddle point $III$ in the system. This is the analogue of the Hagedorn transition at $d \ne 0$.

\begin{figure}[h!]
 \begin{center}
   \psfrag{xlabel}{$\rho_1$}
   \psfrag{ylabel}{$\frac1{N^2} S_\mathrm{tot}(\rho_1)$}
   \psfrag{zlabel}{$\frac1{N^2}\frac{\del S_\mathrm{tot}}{\del \rho_1}$}
   \psfrag{key1}[r][r][0.5]{$T_1 < T_c$}
   \psfrag{key2}[r][r][0.5]{$T_2 > T_c$}
   \psfrag{key3}[r][r][0.5]{$T_3 > T_c$}
   \subfigure[$\frac1{N^2}\frac{\del S_\mathrm{tot}}{\del \rho_1}$]{\label{fig:endof1a} \includegraphics[width=5.5cm]{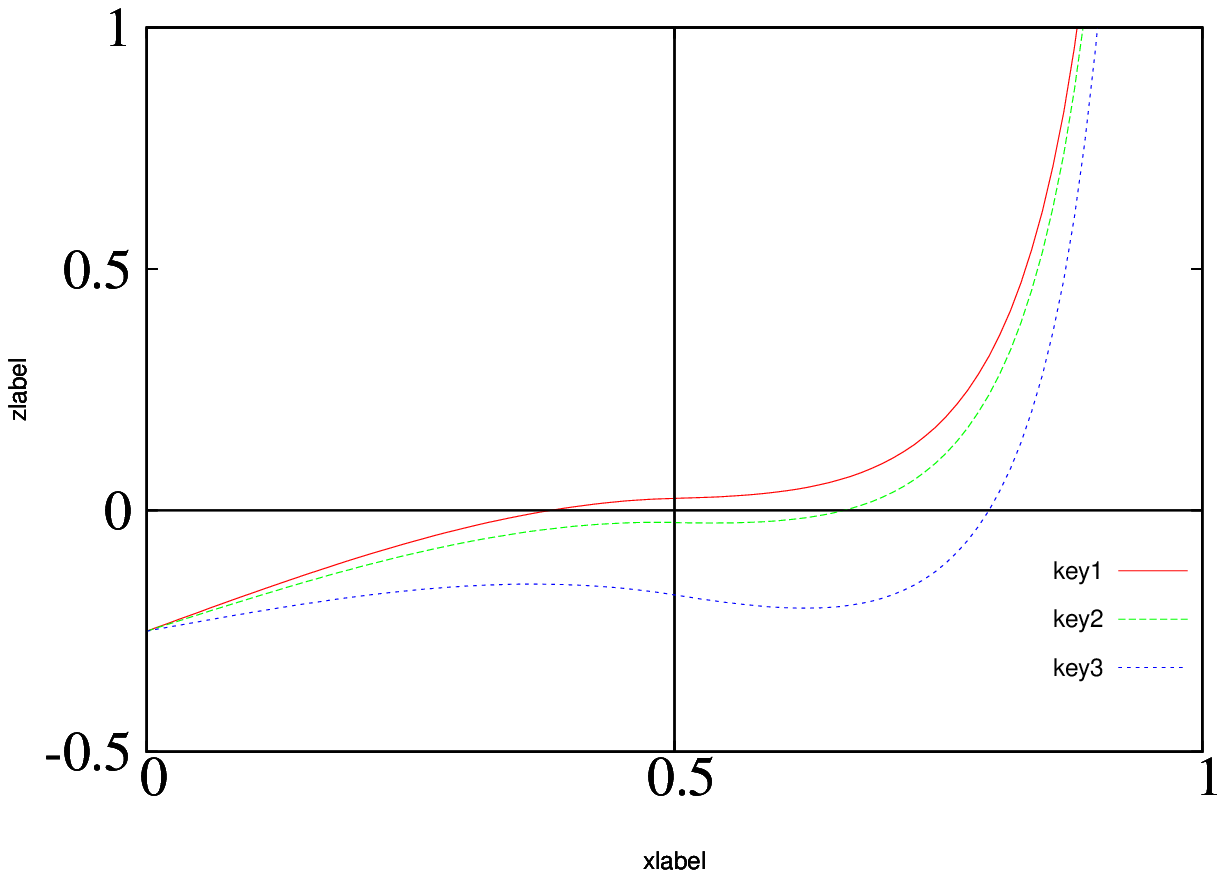}} \hspace{0.1cm}
   \subfigure[$\frac1{N^2} S_\mathrm{tot}(\rho_1)$]{\label{fig:endof1b} \includegraphics[width=5.5cm]{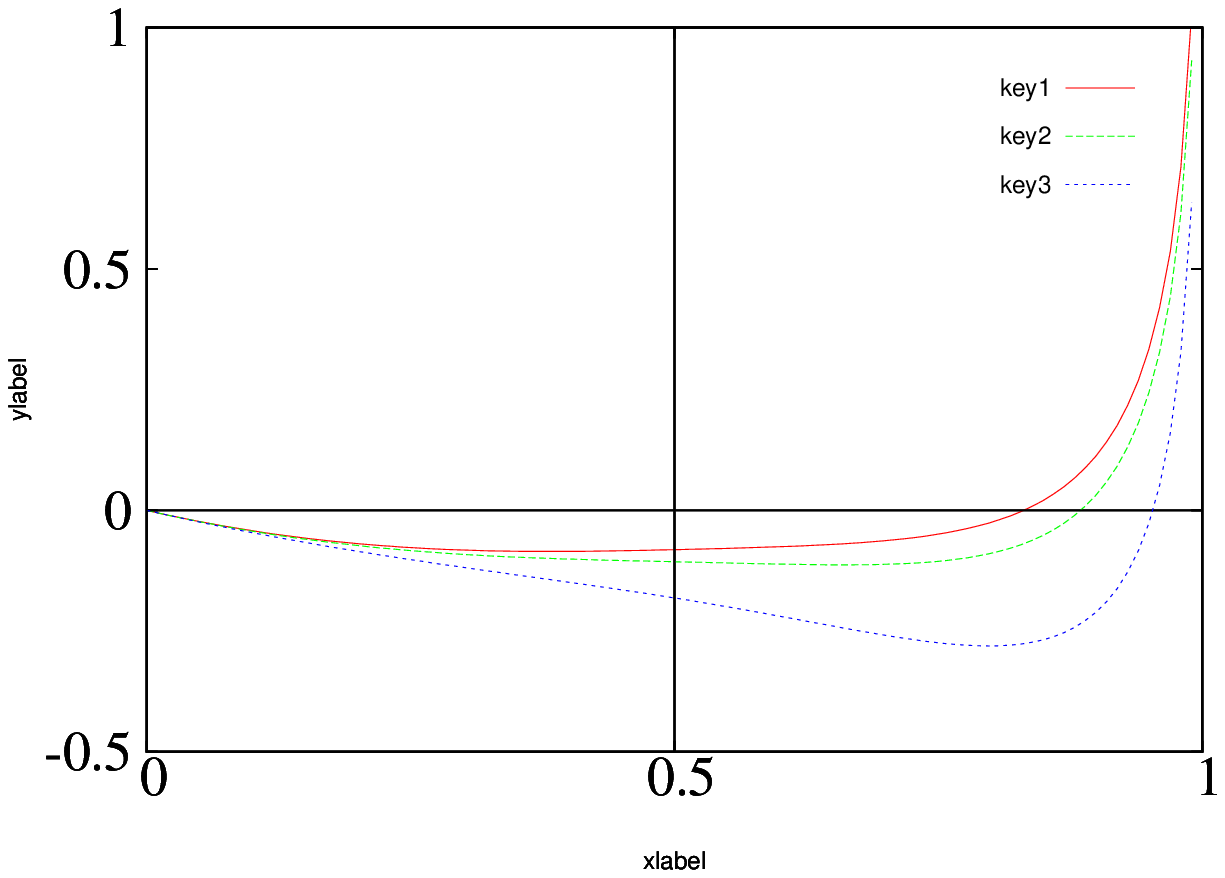}}
 \end{center}
 \caption{End of first order phase transition at $N_f/N \ne 0$.}
 \label{fig:endof1}
\end{figure}

As we increase the value of $d$, the difference between $S_{III}-S_{I}$ at $m_1^{2,\mathrm{1^{st}}}(d)$ (deconfinement temperature) decreases. This is due to the offset created by nonzero $d$ in \eref{saddle2}. Hence there is a critical $d=d_c$ at which all three saddle points of the system meet at a critical temperature $m_1^2 = m_1^{2,c}(d_c)$. If we increase $d$ more than $d_c$ then the new saddle points do not appear any more and the system just has one saddle point. The value of $\rho_1$ in this saddle point increases from $\rho_1=0$ with increasing temperature and there is no deconfinement transition\footnote{The phase diagram here has strong resemblance with the phase diagram of R-charged black holes in $AdS_5$ \cite{Chamblin:1999hg,Basu:2005pj}. Also,in \cite{Headrick:2007ca} the introduction of chemical potential for Maldacena loop gives rise to phase diagram which is quite similar to our model.}. Thus the line of first order transitions ends at a finite $d=d_c$. This is illustrated in Fig \ref{fig:endof1} for $d > d_c$. We see as the temperature is increased there is is no nucleation of new saddle points.

\nt When $d>d_c$, there is a GWW type third order transition for the saddle point as $\rho_1$ crosses $\frac12$. We believe that this should not be interpreted as an analogue of deconfinement transition and should be interpreted along the lines of \cite{AlvarezGaume:2006jg}.

Let us now summarize the resulting phase structure as a function of the three parameters $\lambda, T, N_f/N$ (or equivalently $m_1^2, b, d$). We see that two different transitions are taking place: a first order deconfinement transition which ends on a critical line, and a third order GWW transition which is always present.
\begin{figure}[h!]
 \begin{center}
  \subfigure[$1^{\rm st}$ order surface]{\label{fig:1order} \includegraphics[width=5.5cm]{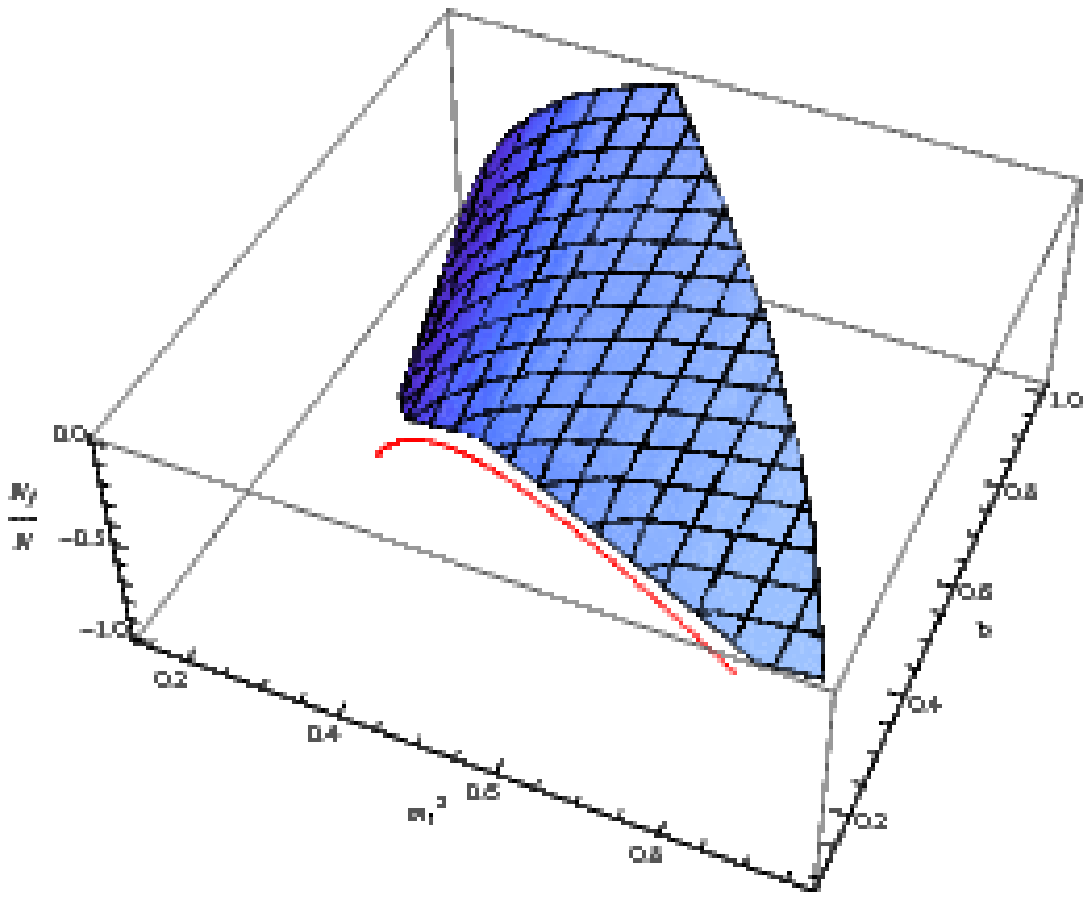}}
  \subfigure[$3^{\rm rd}$ order surface]{\label{fig:3order} \includegraphics[width=5.5cm]{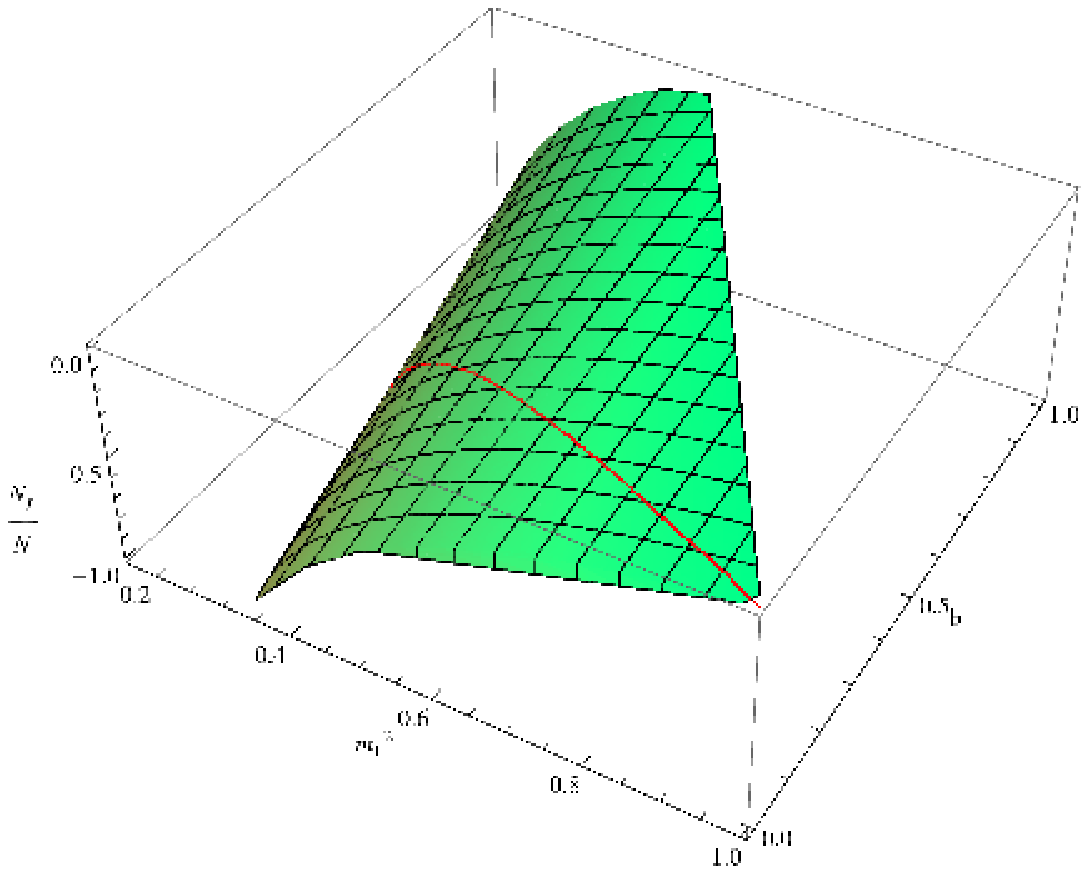}}
 \end{center}
 \caption{Phase diagrams of the system.}
 \label{fig:3dphase}
\end{figure}
In Fig \ref{fig:1order} the red line marks the position of the critical points where the three saddles meet, and the first order phase transition ends. The surface marks the locus of points where saddle $III$ becomes dominant. The first order transition takes place when this surface is crossed. Fig \ref{fig:3order} shows the locus of points where the third order transition takes place\footnote{In the figure, the critical line lies on the surface of third order transition points. This is an artifact of our $\tr\,U^4$ model and not a generic feature. See Appendix \ref{artifact} for more details.}. At the parameter values corresponding to a point on this surface the saddle $II$ crosses $\rho_1 = \frac12$. The nature of the saddle solutions is illustrated in Fig \ref{fig:3dplot}, for $b = \frac12$. The vertical plane is at $\rho_1 = \frac12$. The horizontal plane is a surface of constant temperature. We see that for large enough nonzero $N_f/N$, the saddle profile becomes monotonic. So at any temperature there is only one saddle and there is no first order transition. For small $N_f/N$ the saddle profile is non-monotonic, so above some temperature $T_N$ there are three saddles, and a first order transition takes place. Note that there is always a third order transition at any $N_f/N$, and this happens when the constant temperature surface intersects the saddle point solution surface at $\rho_1 = \frac12$.
\begin{figure}[h!]
 \begin{center}
  \subfigure[$m_1^2 = 0.3$]{\label{fig:3dplot1} \includegraphics[width=5.5cm]{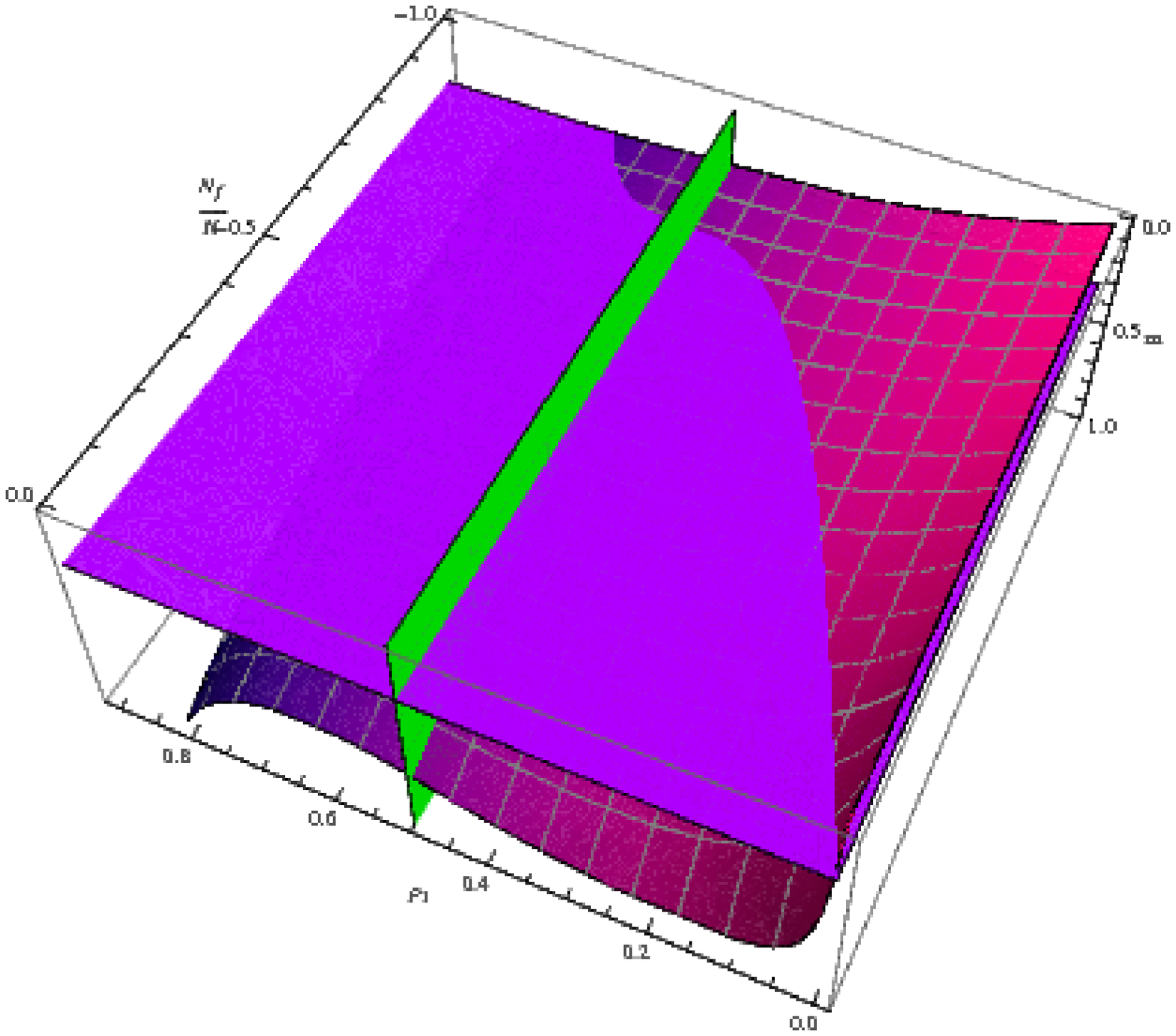}}
  \subfigure[$m_1^2 = 0.6$]{\label{fig:3dplot2} \includegraphics[width=5.5cm]{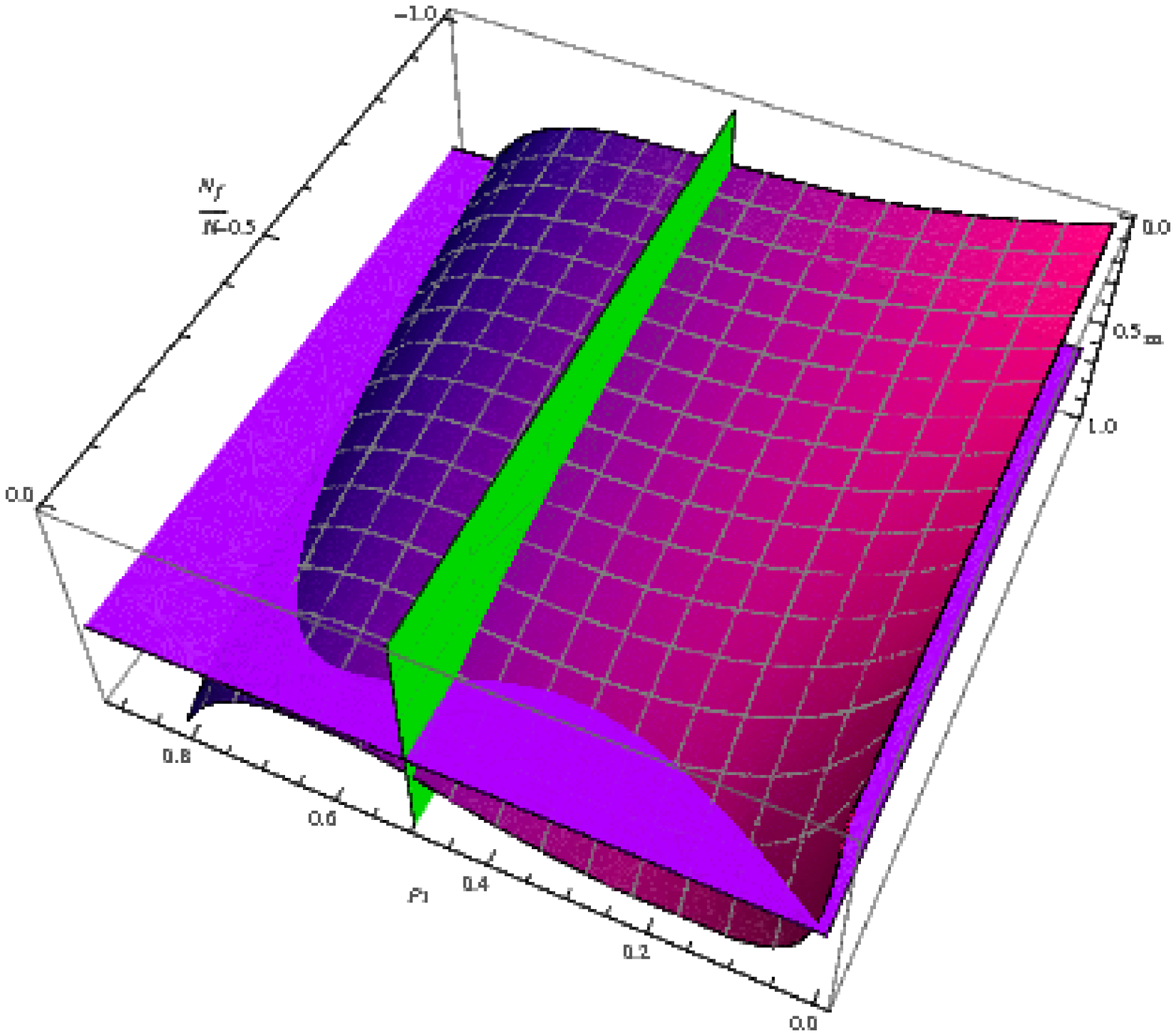}}
 \end{center}
 \caption{Saddle points.}
 \label{fig:3dplot}
\end{figure}
For ease of understanding we present a schematic diagram (Fig \ref{fig:3saddle}) of the saddle point dynamics for a fixed coupling ($\lambda$) and varying $N_f$ and $T$. The bottom most graph (dotted, graph 1) is  where nucleation of saddles $II$ and $III$ take place. The graph (solid, graph 2) above it is the line of first order transitions and the semi-dotted line (graph 3) with arrow is the line of third order transitions. The upper most graph (graph 4) is where the saddle $I$ becomes locally unstable. Three of the  graphs (graph 1,2,4) meet and end at a critical point. The critical point itself lies on the line of third order transitions(i.e. graph 3). As noted above this is not a generic feature. It should be noted that the line of third order transitions becomes physical (in the sense that the dominant saddle point passes through it) after the critical point. 
\begin{figure}[h!]
 \begin{center}
\psfrag{dc}{$d_c$}
\psfrag{T}{$T$}
\psfrag{d}{$d$}
   \includegraphics[width=5.5cm]{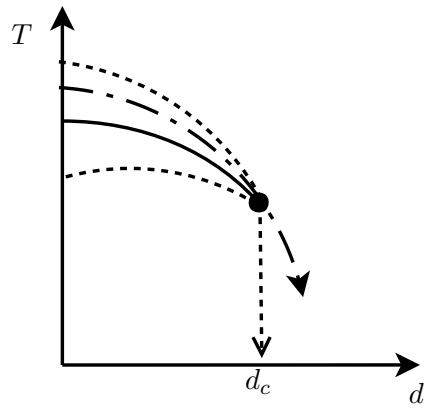}
 \end{center}
 \caption{Saddle dynamics}
 \label{fig:3saddle}
\end{figure}

\section{Discussion and future directions}

In this work we have shown that for a weakly coupled gauge theory on $S^3$, introduction of flavour degrees of freedom makes the deconfinement transition non-existant. It is natural to ask whether we can extrapolate our result to strongly coupled ${\cal N}=4$ SYM theory or even strongly coupled large $N$ QCD on $S^3$. We note that the phase structure of weakly coupled ${\cal N}=4$ SYM theory on $S^3$ is similar to that obtained from strong coupling calculations done using supergravity analysis. It is speculated that this phase structure continues to be valid in the intermediate coupling regime\cite{Aharony:2003sx}. We conjecture that even at finite $N_f/N$, one can extrapolate from a weakly coupled phase diagram to strongly coupled one. Essentially the nature of the phase diagram in Fig \ref{Fig1} remains the same as we change the coupling. This is motivated from the observation that introduction of flavours generically induces a linear term (in $TrU$) in the effective action (\eref{genform}), which one can predict just from gauge invariance. At strong or intermediate coupling, we have little analytic control over the coefficients of such terms. However it is natural to speculate that any linear term will create an offset, which is similar to $d$ in \eref{wthflav} and a mechanism similar to the one in section \ref{1orderterm} may be taking place.

One may also try to study the finite flavour theory from the supergravity side. At zero temperature, the exact back reacted $D3/D7$ brane solution is known. The near horizon geometry of such a solution has an interpretation of SYM theory with $N_f$ fundamental flavours. The finite temperature solution is not known, but we expect that the D7 branes will become non-extremal. Hence even at small temperatures there will be creation of a horizon. This fact tallies with the observation that for finite $N_f$, $\rho_1$ is always non-zero at finite temperature. What happens in the brane picture at higher temperature is less clear. Because of the lack of symmetry, this is a hard problem to solve analytically (D7 branes are localized in the internal manifold $S^5$ and consequently the geometry depends both on the internal coordinates and the radial coordinate). One may hope to make progress in the dissolved brane scenarios \cite{Casero:2006pt,Canoura:2008at}, which makes the solution independent of the radial coordinate and a model similar to \cite{Headrick:2007ca} may emerge (see appendix \ref{phenmodel} for a detailed discussion).

In the case of the pure YM theory, the coupling can be tuned by defining it over the sphere ($S^3$) of varying radius ($R$). The theory will be weakly coupled if $R \ll \Lambda_{YM}$ and strongly coupled in the $R \rightarrow \infty$ limit. Hence the perturbative deconfinement transition in small $S^3$ may indeed be connected to the deconfinement transition in flat space for large $N$ (for finite $N$, deconfinement transition does not happen in finite volume and only happens in flat space). In a similar spirit our finite $N_f/N$ result may also be extrapolated to $R \rightarrow \infty$. The connection with real QCD is more speculative. In real QCD $N=3$ and $N_f \approx 2-3$ and the theory is defined on the flat space. However the non-existence of a sharp first order transition in QCD may indeed be the extrapolation of our result at strong coupling and finite $N$. 

\begin{figure}[h!]
 \begin{center}
\psfrag{infty}{$\infty$}
\psfrag{m1,m2=infty}{$m_1,m_2=\infty$}
\psfrag{m1}{$m_1$}
\psfrag{m2}{$m_2$}
   \includegraphics[width=5.5cm]{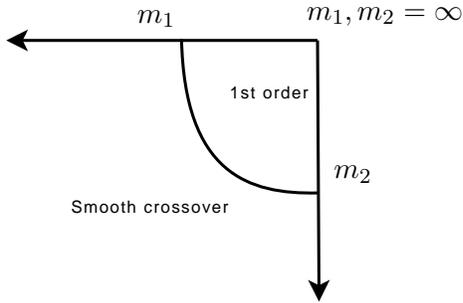}
 \end{center}
 \caption{Diagram showing different regions in parameter space}
 \label{fig:cross}
\end{figure}

As we have commented earlier $d$ in \eref{wthflav} may be varied by varying the mass of the fundamental degrees of freedom and decreasing the mass of the quarks increases $d$. We may compare our results with the phase structure for $YM$ theory with varying quark masses determined from the lattice simulations, as shown in Fig 1 of \cite{Petreczky:2004xs} (and references therein). It is shown that decreasing the quark masses, starting from infinity, changes the degree of the deconfinement transition from first order to second order/smooth crossover. Our result may also be interpreted in a similar manner. The diagram presented in \cite{Petreczky:2004xs} is for $YM$ theory with three quarks. Two of which have the same mass (like the $u,d$ quarks) and the third one has a different mass (like the $s$ quark). We can simulate a similar situation by breaking $N_f$ in two parts of size $2\frac{N_f}{3}$ and $\frac{N_f}{3}$ respectively. We can give these set of quarks different masses $m_1$ and $m_2$. In the $m_1,m_2$ plane a constant $d$ surface(i.e. $d(T_c)=d_c$) will separate regions of first order transition and the regions of smooth crossover. For large $m_1,m_2$,  we can use the relation $d(T) \propto 2\exp({-\frac{m_1}{T}})+\exp({- \frac{m_2}{T}})$ to draw such a surface (Fig \ref{fig:cross}), which has the same generic features as of the diagram presented in \cite{Petreczky:2004xs} \footnote{One should keep in mind that we do not have a issue of chiral symmetry breaking for our problem. All the fields on $S^3$ are massive due to the curvature of $S^3$. Hence the region near $m=0$ is excluded from our diagram.}. 
\par
 Another interesting avenue is to study the  effects of flavours on the lower dimensional gauge theories, possibly using numerical methods. We are planning to address this issue in a future paper \cite{fut}.  
\section*{Acknowledgements}
We thank Mark Van Raamsdonk for many useful comments and suggestions on the draft. We also thank Takehiro Azuma, Joanna Karczmarek, Henry Ling, Sunil Mukhi, Moshe Rozali, Gordon Semenoff, Hsein-Hang Shieh and Jackson Wu for discussions and encouragement. PB thanks Spenta Wadia for useful comments and for suggesting references. We acknowledge support from the Natural Sciences and Engineering Research Council of Canada.
\appendix

\section{An artifact of the $\tr\,U^4$ model}
\label{artifact}

\begin{figure}[h!]
 \begin{center}
\psfrag{dc}{$d_c$}
\psfrag{T}{$T$}
\psfrag{d}{$d$}
   \includegraphics[width=5.5cm]{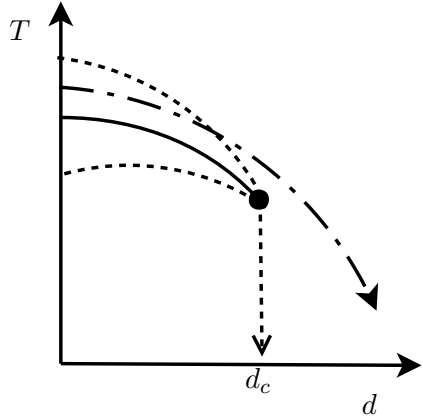}
 \end{center}
 \caption{Saddle dynamics for a generic model}
 \label{fig:3saddle2}
\end{figure}

In this paper we have considered an effective action for the $SU(N)$ theory keeping terms up to $\mathcal{O}(\lambda^2)$ in the double expansion in $\lambda$ and $N_f/N$. From \eref{effective} only the first three terms have coefficients to this order in $\lambda$, so the effective action has a $\rho_1^4$ term as the maximum power of $\rho_1$. If we keep higher orders in $\lambda$ we can have terms with higher powers of $\rho_1$. In the present context we want to check if the coincidence of the critical point and the third order transition which we found earlier continues to be true for models with with higher powers of $\rho_1$. To that end, let us consider a generic model, with
\begin{equation}
 \label{generic}
 N^{-2}S_\mathrm{eff} = -m_1^2 \rho_1^2 - b \rho_1^p + 2 d \rho_1
\end{equation} 
At the critical point three saddle points of the system merge, because the saddle point equation ($S'_{\mathrm{tot}}(\rho_1)$) develops a third order zero. The condition for the critical point is thus
\begin{eqnarray}
 \label{critical}
 \nn S'_\mathrm{tot}(\rho_1)=S'_\mathrm{eff}(\rho_1)+S'_\mathrm{M}(\rho_1)&=&0 \\
 S''_\mathrm{tot}(\rho_1)=S''_\mathrm{eff}(\rho_1)+S''_\mathrm{M}(\rho_1)&=&0 \\
 \nn S'''_\mathrm{tot}(\rho_1)=S'''_\mathrm{eff}(\rho_1)+S'''_\mathrm{M}(\rho_1)&=&0,
\end{eqnarray}
where $S_\mathrm{M}$ is the contribution from the $U(N)$ measure. The first one of the above three conditions is satisfied by changing the parameter $d$. For $\rho_1 \ge \frac12$, the remaining two can be written as:
\begin{eqnarray}
 \label{critical2}
 m_1^2 + 2bp \rho_1^{p-1} &=& \frac{1}{4(1-\rho_1)^2} \\
 \nn b p(p-1) \rho_1^{p-2} &=& \frac{1}{4(1-\rho_1)^3}.
\end{eqnarray}
The solutions to the above equations can be written as:
\begin{eqnarray}
 \label{critical3}
 b &=& \frac{1}{4p(p-1)(1-\rho_1)^3} \, \rho_1^{2-p} \\
 \nn m_1^2 &=& \frac{(p-1)-\rho_1(p+1)}{4(p-1)(1-\rho_1)^3}
\end{eqnarray} 
If we impose the probable physical condition $m_1^2 \ge 0$, then we get
\begin{equation}
 \label{critcal4}
 \frac{p-1}{p+1} \ge \rho_1 \ge \frac12 \\
\end{equation}
In our case $p=3$, so $\rho_1 = \frac12$ always at the critical point. Hence the line of third order transitions goes through $T_\mathrm{crit}$ for our model. This is not true in a generic model with a higher value of $p$. In contrast with Fig \ref{fig:3saddle}, here we have shown in Fig \ref{fig:3saddle2} that the critical point does not necessarily lie on the line of third order transitions. The rest of the features of Fig \ref{fig:3saddle2} are similar to those of Fig \ref{fig:3saddle}.

\section{A phenomenological model at strong coupling}\label{phenmodel}
Although it is difficult to study the supergravity systems at finite temperature with considering the effects of flavour back reaction, we may still motivate a model with a supergravity action which incorporates a linear term of order parameter ($\tr U$) in the effective action. This is done by introducing a chemical potential for the string world sheet in $AdS_5$, which terminates on a Wilson loop in the boundary\footnote{Actually, this introduces a chemical potential for Polykov-Maldacena loop in the boundary. But we will neglect this discrepancy for our phenomenological model.}. This system is discussed in \cite{Headrick:2007ca,Guendelman:1991qb}. To study $S^3$ independent physics, an average is taken over the angular directions in $AdS_5$ and it is assumed that metric depends only on the radial coordinate. Using the radial symmetry, we may write the order parameter as,
\begin{equation}\label{OP1}
\Phi[g] = \Phi_{\rm ct} -T\int\frac{d^p\theta}{\Omega_p}d\tau dr\,(g_{rr}g_{\tau\tau})^{1/2}
= \Phi_{\rm ct} - \int_{r_{\rm min}}^\infty dr\,(g_{rr}g_{\tau\tau})^{1/2}\,.
\end{equation}
Varying the Lagrange multiplier term $-\kappa(\Phi[g]-\Phi_0)/T$ with respect to the metric gives a contribution to the stress tensor which is that of a relativistic string of tension $\kappa$ extended in the $\tau$ and $r$ directions and smeared over the angular directions. The Einstein equation is now: (with $p=3$ for $AdS_5$, as the boundary sphere is a $S^3$)
\begin{equation}\label{Einstein}
{G^\tau}_\tau = {G^r}_r = \frac12p(p+1) - \frac{8\pi G_{\rm N}\kappa}{\Omega_pg_{\Omega\Omega}^{1/2}}\,, \qquad
{G^\theta}_\theta = \frac12p(p+1)\,.
\end{equation}

These equation has a Schwarzschild like solution with metric , 

\begin{equation}\label{Smetric}
ds^2 = f(r)d\tau^2 + f(r)^{-1}dr^2 + r^2d\Omega_p^2\,.
\end{equation}
where,
\begin{equation}\label{fdef}
f(r) = r^2 + 1  - \frac{16\pi G_{\rm N}\kappa}{p\Omega_pr^{p-2}} - \frac\mu{r^{p-1}}\,.
\end{equation}
Here $\mu$ is fixed by requiring the metric to be smooth at the horizon, which sets $f'(r_h)=4\pi T$, where the horizon radius $r_h$ is the largest root of $f$.
In this case $\Phi$ can be calculated easily, as it is independent of the particular form of $f$:
\begin{equation}\label{Phirh}
\Phi = r_h\,.
\end{equation}
The free energy as a function of $\Phi$ may be written as,
\begin{equation}
\frac{16\pi G_{\rm N}F}{\Omega_p} = p\Phi^{p+1} - 4\pi T\Phi^p + p\Phi^{p-1}\,.
\end{equation}
The plot of the free energy is shown in Fig \ref{fig:FS3} (taken from \cite{Headrick:2007ca}):
\begin{figure}[h!]
 \begin{center}
   \includegraphics[width=6cm]{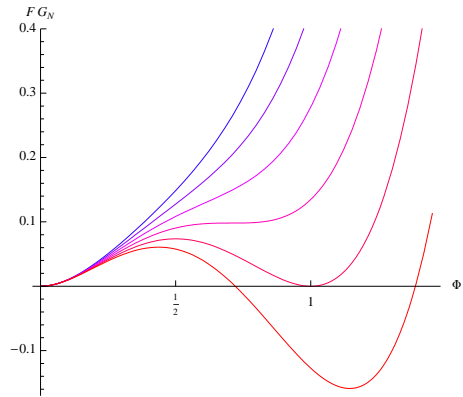}
 \end{center}
 \caption{Saddle dynamics: plot of free energy for various values of T (increasing from the bottom) showing emergence and joining of the saddle points. }
 \label{fig:FS3}
\end{figure}

\nt The corresponding phase diagram is similar to Fig \ref{Fig1} and is shown in  Fig 3 of \cite{Headrick:2007ca}.

\bibliographystyle{hunsrt}
\bibliography{qcd}

\end{document}